\documentclass[12pt,preprint]{aastex}

\slugcomment{To appear in $The$ $Astrophysical$ $Journal$}

\shorttitle{Abundance Pattern in the Fornax dSph Stars}
\shortauthors{Li et al.}

\begin{document}
\title {Investigation of the Puzzling Abundance Pattern in the Stars of the Fornax Dwarf Spheroidal Galaxy}
\author{Hongjie Li\altaffilmark{1,2}, Wenyuan Cui\altaffilmark{1} and Bo Zhang\altaffilmark{1,3}}

\affil{1. Department of Physics, Hebei Normal University, No. 20
East of South 2nd Ring Road, Shijiazhuang 050024, China \\
2. School of Sciences, Hebei University of Science and Technology,
Shijiazhuang 050018, China}

\altaffiltext{3}{Corresponding author. E-mail address:
zhangbo@mail.hebtu.edu.cn}

\begin{abstract}

Many works have found unusual characteristics of elemental
abundances in nearby dwarf galaxies. This implies that there is a
key factor of galactic evolution that is different from that of the
Milky Way (MW). The chemical abundances of the stars in the Fornax
dwarf spheroidal galaxy (Fornax dSph) provide excellent information
for setting constraints on the models of the galactic chemical
evolution. In this work, adopting the five-component approach, we
fit the abundances of the Fornax dSph stars, including $\alpha$
elements, iron group elements and neutron-capture elements. For most
sample stars, the relative contributions from the various processes
to the elemental abundances are not usually in the MW proportions.
We find that the contributions from massive stars to the primary
$\alpha$ elements and iron group elements increase monotonously with
increasing [Fe/H]. This means that the effect of the galactic wind
is not strong enough to halt star formation and the contributions
from massive stars to $\alpha$ elements did not halted for
[Fe/H]$\lesssim$-0.5. The average contributed ratios of various
processes between the dSph stars and the MW stars monotonously
decrease with increasing progenitor mass. This is important evidence
of a bottom-heavy initial mass function (IMF) for the Fonax dSph,
compared to the MW. Considering a bottom-heavy IMF for the dSph, the
observed relations of [$\alpha$/Fe] versus [Fe/H], [iron group/Fe]
versus [Fe/H] and [neutron-capture/Fe] versus [Fe/H] for the dSph
stars can be explained.

\end{abstract}

\keywords{galaxies: abundances--galaxies: dwarf--stars: abundances}

\section{Introduction}

Nearby galaxies can provide many important clues for us to
understand the star formation history, evolution process and
relation between them and the Milky Way (MW). In the past decade,
the accurate abundance of dwarf spheroidal galaxies (dSphs) has
allowed a detailed study of their star formation and chemical
evolution histories \citep{tol03,lan03,lan07,lan08}. Some important
chemical signatures, such as the lower [$\alpha$/Fe] ratios, have
been studied. \cite{der99} found that the galactic winds can remove
a large fraction of gas and the elements produced by massive stars,
which would reduce the enrichment of the interstellar medium (ISM).
\cite{tol03} and \cite{she03} suggested that the low abundance
ratios could be due to a low star formation rate. Based on the study
of chemical evolution of dSphs, \cite{lan03} found that the low
[$\alpha$/Fe] could be explained quantitatively by a sudden decrease
of star formation. Because of relatively shallow potential wells for
dwarf galaxies, the strong galactic winds can remove a large
fraction of the gas and, consequently, the star formation rate drops
to a very low value. In this case, the enrichment of $\alpha$
elements by massive stars is almost halted and SNe Ia are left to
contribute their production, such as iron group elements, due to
longer lifetime. As a result, when the galactic winds occur, the
[$\alpha$/Fe] ratios decrease \citep{lan06,lan07,lan08}.

Using the high resolution spectroscopy, \cite{let10} analyzed the
elemental abundances of the red giant branch stars in the central
region of the Fornax dwarf spheroidal galaxy (Fornax dSph). This
analysis shows the lower ratios of [$\alpha$/Fe], [Ni/Fe] and
[Cr/Fe] compared with those in the MW stars. In contrast, the
abundance of some neutron-capture elements, such as Ba and La, are
enhanced. They found that their sample stars have unusually
abundance patterns and that the Fornax dSph is a chemically complex
system. The Fornax dSph is one of the dwarf spheroidal companions to
the Milky Way. It is a luminous dwarf galaxy with a luminosity of
$M_{v}$= $-13.0\pm0.3$ \citep{irw95} and the second most massive
dwarf galaxy satellites of our Galaxy with a total (dynamical) mass
in the range $10^{8}-10^{9}M_{\odot}$ \citep{mat91,wal06}. Moreover,
it shows a wider range of metallicity and multiple populations
(e.g., \cite{bat06,let10}). The observed abundances of individual
stars could provide an excellent opportunity to study the star
formation history and complicated chemical evolution history of the
Fornax dSph. Recently, \cite{tsu11} studied abundance trends in
Fornax dSph and concluded that a different form of the initial mass
function (IMF) is needed. He found that the lack of very massive
stars (M$\gtrsim25M_{\odot}$) may explain the low [$\alpha$/Fe]
observed in the Fornax dSph. On the other hand, based on the
analysis of the colour-magnitude diagrams and the spectroscopic
metallicity distributions of individual red giant branch stars,
\cite{de12} studied the star formation history of the Fornax dSph.
They found that although the Fornax dSph is dominated by
intermediate age (1-10 Gyr) stellar populations, the star formation
is clearly present at all ages, from 14 Gyr to 0.25 Gyr. This
implies that the galactic wind effect is not intense enough to halt
the star formation in the Fornax dSph.

Elements heavier than iron are mainly synthesized by neutron-capture
process in two ways: one is the slow neutron-capture process (the
s-process) and the other is the rapid neutron-capture process (the
r-process) \citep{bur57}. The s-process is divided into the weak
s-component and the main s-component. The weak s-process occurs in
the massive stars ($M\gtrsim10M_\odot$) undergoing core He burring
and shell C burring \citep{kap89,rai91,rai92,rai93,the00}.
Otherwise, the main s-process elements are produced in the low- to
intermediate-mass stars ($\approx$1.5$-$8$M_\odot$) during the
asymptotic giant branch (AGB) phase \citep{bus99}. Type II
supernovae (SNe II) are usually considered to be the candidates in
which the r-process nucleosynthesis occurs \citep{cow91,sne08}.
Because of the large Eu overabundance ([Eu/Fe]$\sim$1.6), two
metal-poor stars CS 22892-052 and CS 31082-001 arose extensive
attention. These two stars are called as ``main r-process stars",
because their abundance patterns of the heavier neutron-capture
elements match the solar-system r-process pattern well (e.g.,
\cite{cow99,tru02,wan06,sne08}). However, their lighter
neutron-capture elements are too deficient to fit the solar-system's
r-process pattern. This implies that another process, which is
referred to the ``lighter element primary process"
\citep{tra04,cow05,cow06} or ``weak r-process" \citep{ish05}, is
needed. The abundance patterns of weak r-process stars, HD 122563
and HD 88609, show an excess of lighter neutron-capture elements and
a deficiency of heavier neutron-capture elements
\citep{wes00,joh02,hon04}.

Observational abundances of metal-poor stars in the MW show that the
heavier neutron-capture elements and light elements are not produced
in the same sites \citep{joh02,sne03,hon04}. This implies that the
main r-process should occur in O-Ne-Mg core-collapse SNe II with
progenitors of 8-10$M_{\odot}$ \citep{qia07}. Otherwise, the weak
r-process elements are coupled with the light elements and iron
group elements \citep{qia02,qia07,izu09}. The primary light elements
and iron group elements (i.e., the yields independent of initial
metallicity approximately) are produced in massive stars with
$M\gtrsim10M_{\odot}$ \citep{heg10}. \cite{mon07} analyzed the
abundance of metal-poor stars and found that the weak r-process
abundance pattern is uniform and unique. Based on the observed
abundances of weak r-process stars and the main r-process stars,
\cite{li13} derived the abundances of weak r-process and main
r-process using an iterative method and extended ``the weak
r-process component" to primary light elements and primary iron
group elements. They found that the abundances in all metal-poor
stars contain the contributions from the two r-processes.

Because stars of different mass contribute different elements to the
dSph on different timescales, the abundances of individual stars
contain the contributions from various astrophysical processes and
are the integrated results over the lifetime of the system since the
stars were born. In order to reveal the complicated star formation
history and evolution process of Fornax dSph, it is important to
analyze the elemental abundances in detail, including $\alpha$
elements, iron group elements and neutron-capture elements. These
reasons inspired us to study the abundance of the dSph stars. In
this case, the quantitative decomposing stellar elemental abundances
of the contributions from various astrophysical processes are
significant. In this paper, we study the astrophysical origins that
reproduc the abundance pattern of 56 stars in the Fornax dSph, in
which the $\alpha$ elements, iron group elements, lighter
neutron-capture elements and heavier neutron-capture elements have
been observed, and we analyze the relative contributions from the
individual processes. The five-component abundance approach is
described in Section 2. In Section 3, the analysis of calculated
results are presented. Section 4 is our conclusions.

\section{Abundance analysis approach of the stars in the dSph and MW}

Several previous studies on the abundances of neutron-capture
elements have indicated that for most stars, the observed abundances
of the heavy elements cannot be matched by only one neutron-capture
process \citep{tra99,tra04,all06}. One of our major goals is to
explore the astrophysical origin of elements, including light
elements, iron group elements and neutron-capture elements, in the
stars of the Fornax dSph by comparing the observed abundances with
the predicted contributions of various astrophysical processes. The
ith element abundance can be calculated as

\begin{equation}
N_{i}([Fe/H])=(C_{r,m}N_{i,r,m}+C_{pri}N_{i,pri}+C_{s,m}N_{i,s,m}+C_{sec}N_{i,sec}+C_{Ia}N_{i,Ia})\times10^{[Fe/H]}
\end{equation}
where $N_{i,r,m}$, $N_{i,pri}$, $N_{i,s,m}$, $N_{i,sec}$ and
$N_{i,Ia}$ are the abundances of the ith element produced by the
main r-process, the primary process in massive stars, the main
s-process, the secondary process in massive stars, and SNe Ia,
respectively, which have been normalized to abundances of the solar
system; The component coefficients $C_{r,m}$, $C_{pri}$, $C_{s,m}$,
$C_{sec}$ and $C_{Ia}$ represent the relative contributions from the
main r-process, the primary process, the main s-process, the
secondary process and SNe Ia, respectively. Using the five component
coefficients, we can determine the relative contributions of each
process to the elemental abundances and then compare them with the
corresponding component coefficients of the solar system (or
[Fe/H]=0) in which
$C_{r,m}$=$C_{pri}$=$C_{s,m}$=$C_{sec}$=$C_{Ia}$=1. However, these
component coefficients are not usually expected to be equal to each
other. For example, the abundances of neutron-capture elements in
the interstellar gas that formed very metal-deficient stars is
expected to come from mostly r-process events \citep{mon07}. This
method was used by \cite{she13} to study the abundances of Ba stars.
The abundances of $N_{i,Ia}$ are taken from \cite{tim95}, in which
the abundances of Fe, Cu and Zn are adopted from \cite{mis02}.

Observational abundances of metal-poor stars in the MW show that the
weak r-process elements are coupled with the primary light elements
and iron group elements \citep{qia02,qia07,izu09}. Massive stars
(M$\gtrsim10M_{\odot}$) are the astrophysical origins of the primary
elements, including the primary light elements, iron group elements
and weak r-process elements, and provide significant contributions
to the solar abundances for these elements. Note that the mass range
of progenitors in which the light elements and iron group elements
were produced should be different from the progenitor mass range of
SNe II in which the weak r-process elements were produced. The sites
of weak r-process should be the Fe-core-collapse SNe II with
progenitors of mainly about 11-25$M_{\odot}$ \citep{qia07} from
which the primary light elements and iron group elements are also
ejected. However, more massive stars (M$\gtrsim25M_{\odot}$) can
also produce the primary light elements and iron group elements
\citep{heg10}. In this case, the contributions from the more massive
stars should have existed in the observational abundances of the
metal-poor stars and are difficult to separated out. Because the
contributions from the more massive stars, which have shorter
lifetime, are always present at various times, the light elements
and iron group elements in ``the weak r-process component" derived
from the weak r-process stars, HD 122563 and HD 88609, should
contain the contributions from the more massive stars. In this work,
the weak r-process component derived by \cite{li13} is called the
``primary component", since the associated light elements, iron
group elements and the weak r-process elements are produced in
massive stars as primary yields. Note that the best-fitted results
of the metal-poor stars with metallicity range -3.0$<$[Fe/H]$<$-2.0
for the light elements and iron group elements by \cite{li13} imply
that a more complete analysis using the abundances of more
metal-poor stars would likely offer only minor corrections to the
abundances pattern of the primary component. The abundances of main
r-process $N_{i,r,m}$ are adopted from \cite{li13}.

Due to the secondary-like nature of major neutron source
$^{22}Ne(\alpha,n)^{25}Mg$, the contributions from weak s-process
are negligible for metal-poor stars \citep{tra04}. The abundances of
weak s-process are taken from \cite{tra04} at [Fe/H]=0. The ratios
of [$\alpha$/Fe] observed for the MW stars show a plateau at
[Fe/H]$\lesssim$-1, which means that the $\alpha$ elements are
produced mainly by the primary-like mechanisms. Based on
nucleosynthesis calculations, \cite{woo02} found that the effect of
metallicity on the production of $\alpha$ elements in massive stars
is mild. However, the observed increase in some odd-Z elements and
iron group elements toward high [Fe/H] is the effect of metallicity
on nucleosynthesis \citep{mis02,kob06,fel07}. Nucleosynthesis
calculations for massive stars show that the yields of some odd-Z
elements (e.g., Na and Al) and iron group elements (e.g., Mn)
increased with increasing metallicity \citep{woo95,kob06}. The
astrophysical sites of, and dominant contributing processes to, some
iron group elements still have not been reliably established.
\cite{mis02} found that the abundance of iron group element Cu in
the MW contains the contributions from the primary process and
secondary process (i.e., the yields which increase with increasing
initial metallicity) in massive stars. According to estimate of
\cite{mis02}, the contributed fraction from the secondary process in
massive stars for Cu is reached about 25\% for the solar system.
\cite{all11} have found that for disk stars, although the iron group
elements are mostly synthesized by SNe Ia, the secondary-like
contributions from the massive stars are non-negligible.
\cite{she13} found that, for light elements and iron group elements,
the abundances in the solar system are higher than the sum of
contributions from primary component and SNe Ia, because the
contributions from secondary-like yields produced in the massive
stars are not included. In this work, we obtain the secondary-like
abundances for the light elements and iron group elements by
subtracting the sum of the contributions of primary-like yields
(adopted from \cite{li13}), SNe Ia (adopted from \cite{tim95}) and
main s-process component (adopted from \cite{tra04}) from the solar
system abundances (adopted from \cite{and89}). This procedure
ensures that the sum of the abundances of four components is not
overproduced with respect to solar. Because the weak s-process
elements are also produced in massive stars and have a secondary
nature, the secondary-like abundances of light elements and iron
group elements should be tightly related to the weak s-process
abundances.

The secondary light elements, iron group elements and the weak
s-process elements are produced in the massive stars
(M$\gtrsim10M_{\odot}$) \citep{kap89,rai93,the00,mis02,kob06,pig10}.
However, the mass range of the stars in which the light elements and
iron group elements were produced should be different from those of
the stars in which the weak s-process elements were produced. The
sites of the weak s-process (and the part of secondary-like yields
for the light elements and iron group elements) are mainly the
massive stars with $\sim10-30M_{\odot}$ \citep{the07}. However, the
more massive stars (M$>30M_{\odot}$) can also produce the secondary
light elements and iron group elements. The contributions from the
more massive stars are difficult to distinguish in the observed
abundances of the sample stars, because the timescale is long enough
to permit all massive stars to eject their productions for higher
metallicities ([Fe/H]$\gtrsim$-2.0). Recently, \cite{ono12} reported
that the weak s-process also occurs in star with 70$M_{\odot}$. This
would mean that the mass range of stars in which the weak s-process
occurs may be similar to those of stars in which the secondary light
elements and iron group elements were produced. Based on the
discussions above, we combined the secondary-like abundances for the
light elements and iron-group elements with weak s-process
abundances as ``the secondary component". The secondary component
$N_{i,sec}$ contains the abundances of the light elements, iron
group elements and the weak s-process elements.

Based on the observations, \cite{let10} found that compare to the
MW, the ratios of [Ba/Fe] and [Y/Fe] in the dSph show larger
dispersions and [Ba/Y] are exceedingly high. This means that the ISM
in the dSph is chemical inhomogeneous and the abundance pattern of
s-process elements of a newly-formed star was polluted by the winds
of low-mass AGB stars with low-metallicity in the vicinity. In this
case, the s-process nucleosynthesis favors heavier nuclei over
lighter ones, since more neutrons can be captured by an iron seed
nucleus \citep{bus01}. Obviously, there are some differences in the
chemical evolution characteristics between the dSph and the MW,
which lead to the different s-process abundance patterns between the
dSph and the Milky Way. So, the s-process abundance patterns of the
MW are not adequate to used in the dSph stars. Considering the wide
range of metallicity of the sample stars
(-1.4$\lesssim$[Fe/H]$\lesssim$-0.5) in dSph, the abundance pattern
of $N_{i,s,m}$ in equation (1) is calculated as the mixing of the
main s-process abundance produced by low-mass AGB stars with
[Fe/H]=-1.0 and [Fe/H]=-0.6, which are given by \cite{bus01}. We
calculate the main s-process abundance using the equation:
\begin{equation}
C_{s,m}N_{i,s,m}=C_{1}N_{i,s,m}([Fe/H]=-1)+C_{2}N_{i,s,m}([Fe/H]=-0.6)
\end{equation}
in which $N_{i,s,m}$, $N_{i, s,m}$([Fe/H]=-1) and $N_{i, s,m}$
([Fe/H]=-0.6) have been normalized to the main s-process component
of Ba in solar system \citep{arl99}. In this case,
$C_{s,m}$=$C_{1}$+$C_{2}$. By comparison, for the stars in the MW,
the adopted metallicity-dependent abundance pattern $N_{i,s,m}$ in
equation (1) is taken from the main s-process abundance given by
\cite{tra99} (see their Fig. 6-Fig. 12) and \cite{ser09} (see their
Fig. 3-Fig. 6), which included the contributions of low- to
intermediate-mass AGB stars ($\sim1.5-8M_{\odot}$) and also has been
normalized to the main s-process abundance of Ba in solar system.

In equation (1), we have defined five component coefficients. We can
obtain them by looking for the minimum $\chi^2$. The reduced
$\chi^2$ is defined as
\begin{equation}
\chi^2=\sum_{i=1}^K\frac{(logN_{i,obs}-logN_{i,cal})^2}{(\Delta
logN_{i,obs})^2(K-K_{free})}
\end{equation}
where $N_{i,cal}$ is the abundance calculated from equation (1),
$logN_{i,obs}$ and $\Delta logN_{i,obs}$ are the observed
abundance and error of the ith element, respectively, $K$ is the
number of elements applied in the fit and $K_{free}$ is the number
of free parameters. Here, $K_{free}$=5, because equation (1)
contains five component coefficients. Using component
coefficients, we can determine the relative contributions of each
individual process to the stellar abundances. Moreover, we can
compare the derived component coefficients with the corresponding
coefficients of the solar system in which all of them are equal to
1. For the sample stars, the five component coefficients are not
equal to each other, because their elemental abundances are not in
solar proportions.

\section{Results and Discussion}

In this work we analyze the observational constraints provided by
the elemental abundances of the stars in the Fornax dSph to
investigate the relative contributions from the individual process.
Using the observed data of Fornax dSph stars \citep{let10}, the
fitted parameters can be obtained. The derived component
coefficients and $\chi^2$ are listed in Table 1. A comparison
between the abundance-decomposed results of Fornax dSph stars and
the MW stars with a similar metallicity range is necessary to
ascertain whether the elements in the two galaxies have a similar
origin or a different chemical evolution history. Using the observed
data of the MW stars \citep{mis01,red03,red06}, the derived
component coefficients and $\chi^2$ are listed in Table 2. The
[Fe/H] distributions of the dSph stars and the MW sample stars are
compared in Figure 1. One can see that the dSph stars have a similar
distribution of [Fe/H] to that of the Milky Way.

As two examples, the calculated best-fitting results for the dSph
star BL185 and the MW star HIP 72803 are shown in Fig. 2. The solid
lines and filled circles represent the calculated results and the
observed abundances, respectively. The top panels of Figs. 3(a) and
3(b) show the individual relative offsets
($\Delta\log\varepsilon$=$\log\varepsilon_{cal}
-\log\varepsilon_{obs}$) for the dSph stars and MW stars,
respectively. The bottom panels of Figs. 3(a) and 3(b) show the
root-mean-square offset in $\log\varepsilon$. Typical observational
uncertainties in $\log\varepsilon$ are $\sim0.2-0.3$ dex (dash
lines). It could be found from Fig. 3 that the individual relative
offsets are mostly smaller than 0.30 dex and the root-mean-square
offset is consistent with zero. The calculated results shown in
Figs. 2 and 3 confirm the validity of the abundance approach adopted
in this work.

In the study of the abundances of the dSph stars, the component
coefficient as the function of metallicity is very important, since
it shows the chemical evolution features of the dSph. If $C_{k}>1$
(or $C_{k}<1$, k=r,m; pri; s,m; sec and Ia), the contribution from
the process corresponding to the elemental abundances is larger (or
less) than that in the solar system, after excluding the effect of
metallicity. These component coefficients are not usually expected
to be equal to each other, because the relative contributions of
these components are not always in proportion to those of the
solar-system. The left panel and right panel in Fig. 4 show the
component coefficients $C_{r,m}$, $C_{pri}$, $C_{s,m}$, $C_{sec}$
and $C_{Ia}$ as a function of [Fe/H] for the dSph stars and the MW
stars, respectively. In this work, the adopted primary component and
secondary component contain the abundances of light elements, iron
group elements and neutron-capture elements. For neutron-capture
elements, $C_{pri}$ and $C_{sec}$ represent the component
coefficients of weak r- and weak s-process, respectively. For light
elements and iron group elements, $C_{pri}$ and $C_{sec}$ represent
the contributions from primary-like and secondary-like yields in
massive stars, respectively.

For detailed comparisons, Figs. 5 (a)-5 (e) show the distributions
of five component coefficients obtained for the dSph stars (solid
lines) and the MW stars (dash lines). There is a tendency for the
component coefficients $C_{r,m}$, $C_{s,m}$ and $C_{Ia}$ of the dSph
stars to be larger than those of the MW sample stars. This means
that the contributions from main r-process, main s-process and SNe
Ia to the abundances of the dSph stars are larger than those of the
MW stars. However, a trend exists wherein the component coefficients
$C_{pri}$ and $C_{sec}$ of the dSph stars are smaller than those of
the MW stars. This means that the contributions from the primary
process (or weak r-process) and the secondary process (or weak
s-process) to the dSph stars are smaller than those to the MW stars.

The stellar initial mass function (IMF) is an important function of
a galaxy, which can be applied extensively to various aspects of
astrophysics. The IMF of a dwarf galaxy may depend on its dynamics
and kinematics characteristics \citep{dut12} and directly effect the
chemical abundances of the dwarf galaxy stars. Because the $\alpha$
elements, iron group elements, lighter neutron-capture elements and
heavier neutron-capture elements have different astrophysical
origins, one could investigate the IMF using the contributions of
various processes to the abundances of the dSph stars. Recall that
the main r-process should occur in O-Ne-Mg core-collapse SNe II with
progenitors of 8-10$M_{\odot}$ and the primary elements (including
primary light elements, iron group elements and weak r-elements) are
produced in the massive stars with $M\gtrsim10M_{\odot}$. On the
other hand, the secondary elements (including secondary light
elements, iron group elements and weak s-elements) are produced in
the massive stars (M$\gtrsim10M_{\odot}$) and the main s-process
elements are mainly produced in the low-mass AGB stars
($\sim$1.5-3$M_\odot$). In Fig. 6, we show the ratios of the average
component coefficient between the dSph stars and the MW stars as
function of progenitor mass. The ratios monotonously decrease with
the increasing progenitor mass. This means that, compare to the MW,
the bottom-heavy IMF for the Fonax dSph is strongly suggested.

Because stars of various mass contribute different elements to the
dSph on different timescales, the observed abundances of the sample
stars are the integrated results over the lifetime of the system
since the stars were born. In order to make a progress in
understanding the chemical evolution of Fornax dSph in greater
detail, we derive the component ratios of the individual process
(i.e., [element/H]$_{k}$ (k=r,m; pri; s,m; sec; SNe Ia)) with
various metallicities and compare them with those of the MW stars.
In Figs. 7-11, we show the component ratios of Fe, $\alpha$ element
Mg, iron group element Ni, lighter neutron-capture element Y and
heavier neutron-capture element Ba. For ease of comparison with the
solar system, we add some lines in these figures. The dash dotted
lines, solid lines, short dotted lines, dash lines, and dotted lines
represent the solar component ratios of the main r-process, primary
process, main s-process, secondary process and SNe Ia, respectively.
From Fig. 7 (b), we can see that for the MW stars, the contributions
from the primary-like yields to [Fe/H] ratios are larger than those
from SNe Ia and the secondary-like yields. However, Fig. 7 (a) shows
that the Fe abundances for the dSph stars predominantly come from
SNe Ia and the ratios of SNe Ia have reached the solar ratio. These
results are consistent with the bottom-heavy IMF for the dSph. In
this case, the contributions from the primary-like yields and
secondary-like yields produced in the massive stars to the dSph are
smaller than corresponding contributions to the Milky Way.

From Fig. 8 (b) we can see that, for the MW stars, the Mg abundances
predominantly come from the primary process and the ratios are
larger than the solar ratio. In Fig. 8 (a), we find that the Mg
abundances in the dSph stars also mainly come from the primary
process in the massive stars. However, the ratios of the primary
process of the dSph stars are close to that of the solar system and
lower than those of the MW stars. From Fig. 8 (a) we can see that
the ratios of [Mg/H]$_{pri}$, which represent the contributions from
massive stars to the primary yields, increase monotonously with
increasing [Fe/H]. This means that the massive stars have continued
chemical enrichment during the dSph evolution, even though they do
not dominate the Fe enrichment. In other words, for the Fornax dSph,
the stars formed at all ages and the contributions from the massive
stars to $\alpha$ elements did not stop, at least for [Fe/H]
reaching to -0.5. The continual contributions from the massive stars
to the primary yields of iron group elements for the metallicity
range from [Fe/H]=-1.0 to [Fe/H]=-0.5 can also be found from Figure
7 and Figure 9 for iron group elements Fe and Ni. The correlation
between [Mg/H]$_{pri}$ and [Fe/H] obtained from this work is a
significant evidence that the effect of the galactic wind is not
strong enough to halt the star formation in the Fornax dSph. Because
of the bottom-heavy IMF for the dSph, the contributions from the
primary-like yields produced in the massive stars to the dSph are
smaller than corresponding contributions to the Milky Way.
\cite{let10} have found that the observed [Mg/Fe] of the dSph stars
are lower than those of the MW stars and close to the solar ratio
(see their Fig. 10). After considering the bottom-heavy IMF for the
dSph, the abundance ratios of $\alpha$ elements (e.g., Mg, Si, Ca
and Ti) observed in the dSph stars can be explained.

In Fig. 9 (b), for the MW stars, the calculated results imply that
the Ni abundances mainly come from the contributions of the primary
process and secondary process. The component ratios of the primary
process are larger than that of the solar primary ratio, and the
component ratios of the secondary process have reached the solar
secondary ratio at [Fe/H]$\sim-0.8$. Although the component ratios
of SNe Ia are smaller than the solar ratio, the component ratios of
the primary- and the secondary-like yields produced in massive stars
are larger than the corresponding solar ratios. These are the
reasons for the observed [Ni/Fe]$\approx$0 (see Fig. 12 in
\cite{let10}) in the MW stars. In contrast to the Milky Way, from
Fig. 9 (a) we can see that for the dSph stars, the component ratios
of SNe Ia for Ni are larger than those of the primary process and
the secondary process. The ratios of the SNe Ia are close to the
solar ratio of the SNe Ia, and the ratios of the secondary process
are lower than the solar secondary ratio by about 0.5 dex.
\cite{let10} have found that the observed [Ni/Fe] of the dSph stars
are lower than those of the MW stars and the solar system (see their
Fig. 12). These results are consistent with a bottom-heavy IMF for
the dSph. In this case, although the component ratios of SNe Ia have
reached the solar ratio, the component ratios of the primary process
and the secondary process for the dSph are smaller than the
corresponding ratios of the Milky Way. These are the reasons of
observed [Ni/Fe]$<$0 for the Fornax dSph stars. The abundance ratios
of another iron group element Cr observed in the dSph stars can also
be explained after considering the bottom-heavy IMF.

From Fig. 10(b), we can see that for the MW stars, the Y abundances
mainly come from the weak r-process and main s-process. The
component ratios of the weak r-process are larger than the solar
ratio of weak r-process and the component ratios of the main
s-process are smaller than the solar ratio of main s-process. On the
other hand, from Fig. 10 (a) we can see that, for most dSph stars,
contributions from the main s-process exceed main r-process and weak
r-process contributions, and the weak s-process contributions are
negligible at all metallicities. The ratios of the main s-process
have reached the solar ratio at [Fe/H]$\gtrsim-0.9$. Because of the
bottom-heavy IMF, the contributions from main s-process in the AGB
stars to the dSph are larger than the corresponding contributions to
the MW. Although the average value of [Y/Fe] for dSph stars is close
to that of the MW, the proportions contributed from main s-process,
main r-process and weak r-process are obviously different.

From Fig. 11(b) we can see that for the MW stars, the Ba abundances
mainly come from the main r-process and main s-process. The
component ratios of the main r-process are larger than the solar
ratio of main r-process and the ratios of the main s-process are
slightly lower than the solar ratio of the main s-process. From Fig.
11(a) we can see that, for most dSph stars, contributions from the
main s-process exceed contributions from main r-process for
[Fe/H]$>$-1.0 and begin to drive [Ba/H] upward to [Ba/H]=0. In this
case, the ratios of the main s-process for Fornax stars are larger
than those of the solar system and the MW stars. \cite{let10} have
found that the observed [Ba/Fe] of most dSph stars are higher than
those of the MW stars and the solar system (see their Fig. 14). It
is natural result that the contributions of main s-process produced
in AGB stars to the dSph are larger than corresponding contributions
to the Milky Way due to the bottom-heavy IMF of the dSph. Similar to
Ba, the La abundance ratios of the dSph stars can be explained.
\cite{let10} have found that [Eu/Fe] of dSph stars are at the same
levels observed in the MW stars for the same [Fe/H]. Our calculated
results about main r-process component coefficients are consistent
with their findings.

The apparent rise of [La/Fe] or [Ba/Fe] observed in Fornax dSph or
Sagittarius dSph is due to pollution by AGB stellar wind. The
calculated results presented by \cite{lan08} show that [La/Fe] (or
[Ba/Fe]) decrease slightly at high metallicities, which is due to
the effects of galactic winds on the star formation rate. This means
that there is a discrepancy between predictions of the chemical
evolution model and observations. Furthermore, \cite{lan08} have
found that an increase in the yields of La in AGB stars by a factor
that is smaller than four is not enough to match the observed
ratios. However, they reported that if the s-process yields of the
AGB stars were increased considerably, the yields could not be used
in the MW and the solar system correctly. Obviously, there is other
reason to explain the rise of [Ba/Fe] or [La/Fe] observed in the
dSphs.

Historically, variations in the IMF have been proposed to reduce the
chemical enrichment from massive stars \citep{mat83}. The truncated
IMF has been used as a possible cause of lowering [$\alpha$/Fe] in
dwarfs \citep{tol03}. Because the s-process yields should mainly be
dependent on the initial mass and metallicity of the low mass AGB
stars, the average yields of the AGB stars with the same initial
mass and metallicity as the dSph should be similar to those of the
MW. In this case, the rise of [Ba/Fe] or [La/Fe] observed in the
dSphs should be attributed to contributions from the larger number
of AGB stars. So, the abundances of s-process elements, Ba and La,
in the dSph can be used to constrain the number of AGB stars that
have polluted the gas in which the star formed and, thus constrain
the IMF. It is interesting to note that, based on our calculation,
the ratio of the average main s-process coefficient between the
Fornax dSph and the MW is about 5.5, which is close to the increased
factor of the AGB yields reported by \cite{lan08}. Our calculated
results mean that the reason for the rise of [Ba/Fe] or [La/Fe]
observed in the dSphs is the bottom-heavy IMF, which favors a larger
number of low-mass AGB stars.

\section{Conclusions}

In dSphs, nearly all chemical evolution and nucleosynthetic
information is in the elemental abundances of stars with various
metallicities. In this work, we investigate the astrophysical
origins of the observed abundances of Fornax dSph stars and derive
the relative contributions from the individual astrophysical
process. Our results can be summarized as follows:

1. Adopting the five-component approach, the abundances of most
sample stars, including $\alpha$ elements, iron group elements,
lighter and heavier neutron-capture elements, can be fitted. The
component coefficients of the main r-process, main s-process and
secondary process of most dSph stars are different from those of the
solar system. However, the component coefficients of primary process
and SNe Ia are close to those of the solar system.

2. The component coefficients of the main s-process, secondary
process, primary process and SNe Ia of the dSph stars are obviously
different from those of MW stars with similar range of metallicity.
The component coefficients of main r-process, main s-process and SNe
Ia of most dSph stars are larger than those of the MW stars. This
means that the contributions from main r-process, main s-process and
SNe Ia to the dSph stars are larger than those to the MW stars.
However, there are the trends for the dSph stars with smaller
$C_{r,w}$, and $C_{s,w}$. This implies that for the dSph stars, the
contributions from primary process and secondary process are smaller
than the corresponding contribution for the MW stars.

3. The detailed abundance trends of five components for Fornax dSph
could provide the information on the chemical enrichment history. We
find that the contribution from massive stars as the primary yields
of $\alpha$ elements and iron group elements increases monotonously
with increasing [Fe/H]. This means that the star formed, including
massive stars, at all ages and the contributions from the massive
stars to $\alpha$ elements did not halted, at least until [Fe/H]
reached -0.5. These correlations provide the significant evidence
that the effect of the galactic wind is not strong enough to halt
the star formation in Fornax for [Fe/H]$\lesssim$-0.5.

4. The ratios of average component coefficient between the dSph
stars and the MW stars monotonously decrease with the increasing
progenitor mass. The correlation is significant evidence of the
bottom-heavy IMF for the Fonax dSph, compared to the Milky Way.

5. Although Fornax dSph stars have a range of metallicity similar to
the MW sample stars, the contribution proportions for Fe from
various astrophysical processes of Fornax dSph stars are different
with those of the MW stars. For the MW stars the contributions of
the primary-like yields to Fe are larger than those of SNe Ia and
the secondary-like yields. However, the Fe abundances in the dSph
stars predominantly come from SNe Ia. This means that the
contributions from the massive stars to the dSph are smaller than
corresponding contributions to the MW, which is consistent with the
bottom-heavy IMF for the dSph.

6. Although Mg abundances predominantly come from the contributions
of the primary process for both the MW stars and the Fornax dSph
stars, the ratios of the primary process of the dSph stars are close
to that of the solar system and lower than those of the MW stars.
This suggests that the contributions of the primary-like yields
produced in the massive stars to the dSph are smaller than the
corresponding contributions to the MW. After considering the
bottom-heavy IMF for the dSph, the lower [$\alpha$/Fe] observed in
the dSph stars can be explained.

7. For the MW stars, the abundances of the iron group element Ni
mainly come from the primary process and secondary process. However,
for the dSph stars, the component ratios of SNe Ia for Ni are larger
than those of the primary process and the secondary process.
Although the component ratios of SNe Ia has reached the solar system
ratio, the component ratios of the primary process and the secondary
process for the dSph are smaller than corresponding ratios of the
MW. These are the reasons for observed [Ni/Fe]$<$0 for Fornax dSph
stars and are consistent with the bottom-heavy IMF.

8. For the MW stars, the abundances of lighter neutron-capture
element Y mainly come from the weak r-process and main s-process. On
the other hand, for most dSph stars, the contributions from the main
s-process exceed the contributions from main r-process and weak
r-process. The contributions from weak s-process can be negligible
at all metallicities. Although the average value of [Y/Fe] for dSph
stars is close to that of the MW stars, the contributed proportions
of the main s-process, main r-process and weak r-process are
obviously different with those of the MW stars.

9. Because there are some differences in the s-process abundance
characteristics between the dSph and the MW, the s-process abundance
patterns of the MW are not adequate to used in the dSph stars. By
adopting the mixing of the main s-process abundance produced by
low-mass AGB stars with [Fe/H]=-1.0 and [Fe/H]=-0.6 presented by
\cite{bus01}, the abundances of the s-process elements, Ba and Y,
can be fitted. The fitted results mean that the s-process elements
in the dSph predominantly come from the low-mass AGB stars, which is
consistent with the suggestion of the bottom-heavy IMF for the
Fornax dSph.

10. For the MW stars, the abundances of the heavier neutron-capture
element Ba mainly come from the contributions of the main r-process
and main s-process. On the other hand, for the most dSph stars,
contributions from the main s-process to Ba exceed contributions
from the main r-process for [Fe/H]$>$-1.0 and begin to drive [Ba/H]
upward to [Ba/H]=0. The rise of [Ba/Fe] or [La/Fe] observed in
Fornax dSph can be attributed to contributions from the larger
number of AGB stars. It is a natural result that the contributions
of main s-process produced in the low-mass AGB stars to the dSph are
larger than corresponding contributions to the Milky Way due to the
bottom-heavy IMF of the dSph.

Our abundance approach is based on the abundance patterns of the
main r-process, main s-process, SNe Ia, primary process and
secondary process to decompose stellar elemental abundances,
rather than the traditional galactic chemical evolution model
which integrates stellar yields beginning from early galaxies.
This procedure allows us to use available elemental abundances to
investigate their astrophysical origins and chemical evolution
when each contribution is showed as a function of metallicity. We
hope that the results here will provide a useful guide for more
complete chemical evolution models of Fonax dSph. Obviously, a
more precise knowledge about the elemental abundances in the dSph
stars is needed.

\acknowledgments

This work has been supported by the National Natural Science
Foundation of China under Grant No. 11273011, U1231119, 10973006
and 11003002, the Science Foundation of Hebei Normal University
under Grant No. L2009Z04, the Natural Science Foundation of Hebei
Province under Grant No. A2009000251, A2011205102, Science and
Technology Supporting Project of Hebei Province under Grant No.
12211013D and the Program for Excellent Innovative Talents in
University of Hebei Province under Grant No. CPRC034.

\begin{figure}[t]
 \centering
 \includegraphics[width=1\textwidth,height=0.6\textheight]{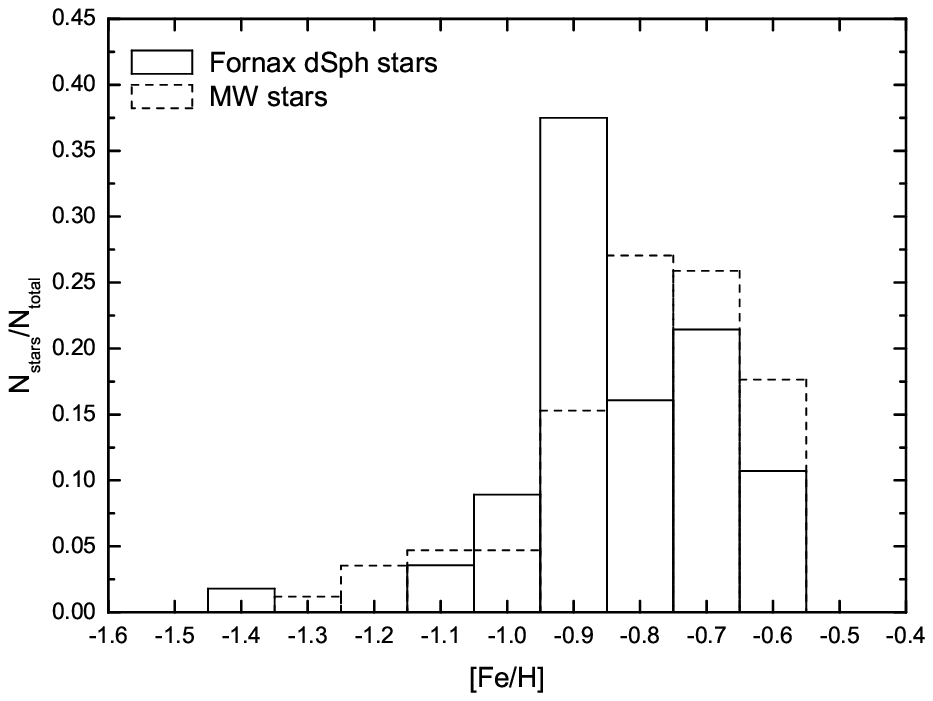}
 %\suppressfloats[t]
\caption{Metallicity distribution in the Fornax dSph stars (solid
lines) compared to that of the MW stars (dash lines). }
 %\label{appenfig}
\end{figure}

\begin{figure}[t]
 \centering
 \includegraphics[width=1.0\textwidth,height=0.4\textheight]{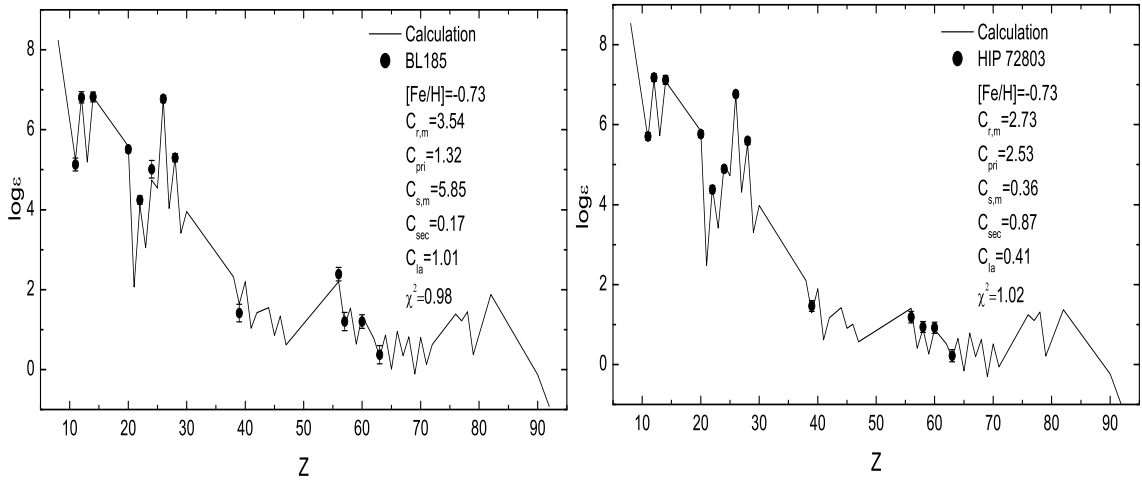}
% \suppressfloats[t]
\begin{center}
\caption{Two examples of calculated best-fitting results for the
sample stars. The solid lines represent the calculated results. The
observed elemental abundances are marked by filled circles. The left
panel is one of the Fornax dSph stars. The right panel is a MW
star.}
% %\label{appenfig}
\end{center}
\end{figure}

\begin{figure}
 \centering
 \includegraphics[width=1.0\textwidth,height=0.4\textheight]{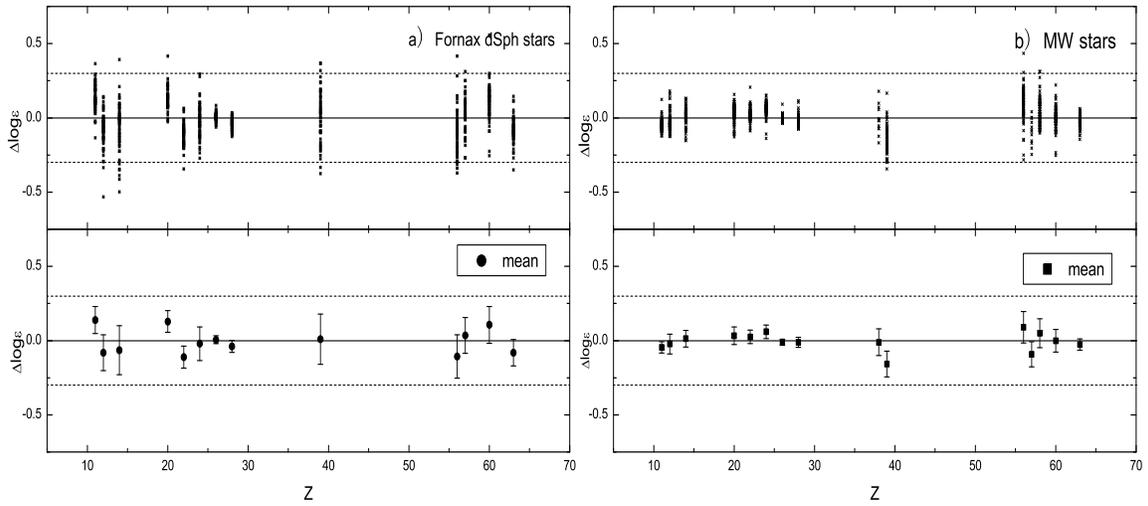}
 %\suppressfloats[t]
\caption{Top panels: individual relative offsets
($\Delta\log\varepsilon(X)$) for the Fornax dSph stars (stars) and
MW stars (crosses). Typical observational uncertainties in
$\log\varepsilon$ are $\sim0.2-0.3$ dex (dash lines). Bottom panels:
The root-mean-square offsets in $\log\varepsilon$ for the Fornax
dSph stars (filled circles) and MW stars (filled squares).}
 %\label{appenfig}
\end{figure}

\begin{figure}
 \centering
 \includegraphics[width=1.0\textwidth,height=0.4\textheight]{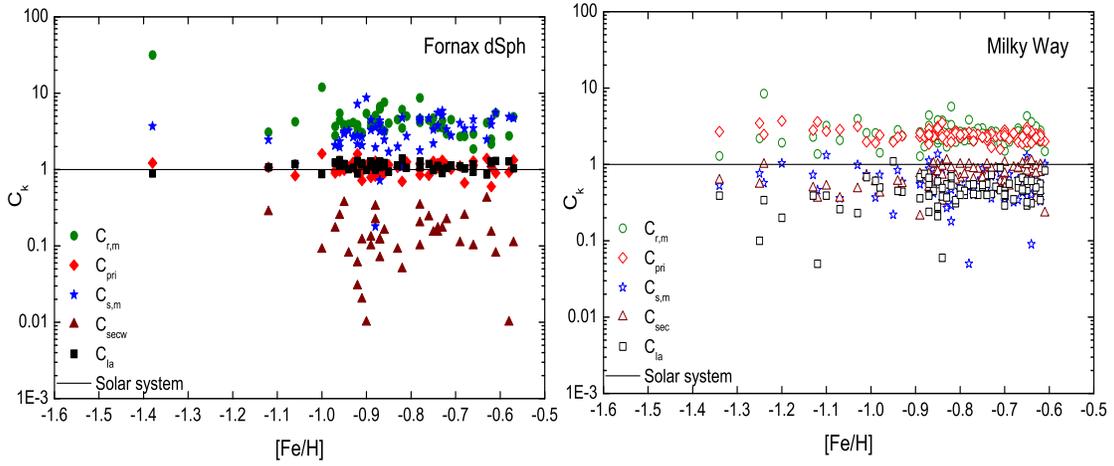}
 %\suppressfloats[t]
\caption{The component coefficients as a function of metallicity. In
the left panel, filed circles, filled diamonds, filled stars, filled
triangles and filled squares are the component coefficients of the
maim r-process, primary process, the main s-process, secondary
process, and SNe Ia component respectively, for Fornax dSph stars.
In the right panel, open circles, open diamonds, open stars, open
triangles and open squares represent the component coefficients of
the maim r-process, primary process, the main s-process, secondary
process, and SNe Ia component, respectively, for MW stars. The solid
line presents the component coefficients of the solar system.}
 %\label{appenfig}
\end{figure}

\begin{figure}[t]
 \centering
 \includegraphics[width=1\textwidth,height=0.6\textheight]{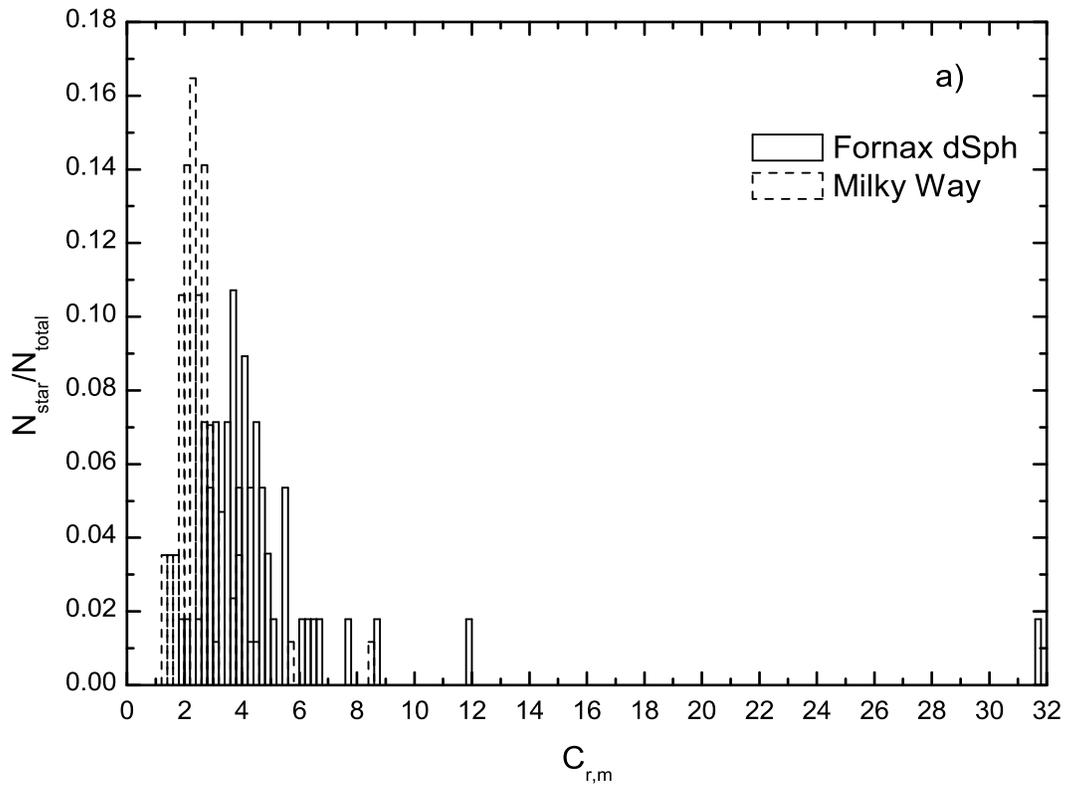}
 %\suppressfloats[t]
\caption{Five component coefficient distributions. The solid lines
represent the Fornax dSph stars. The dash lines represent the MW
stars.}
 %\label{appenfig}
\end{figure}

\begin{figure}[t]
 \centering
 \includegraphics[width=1\textwidth,height=0.6\textheight]{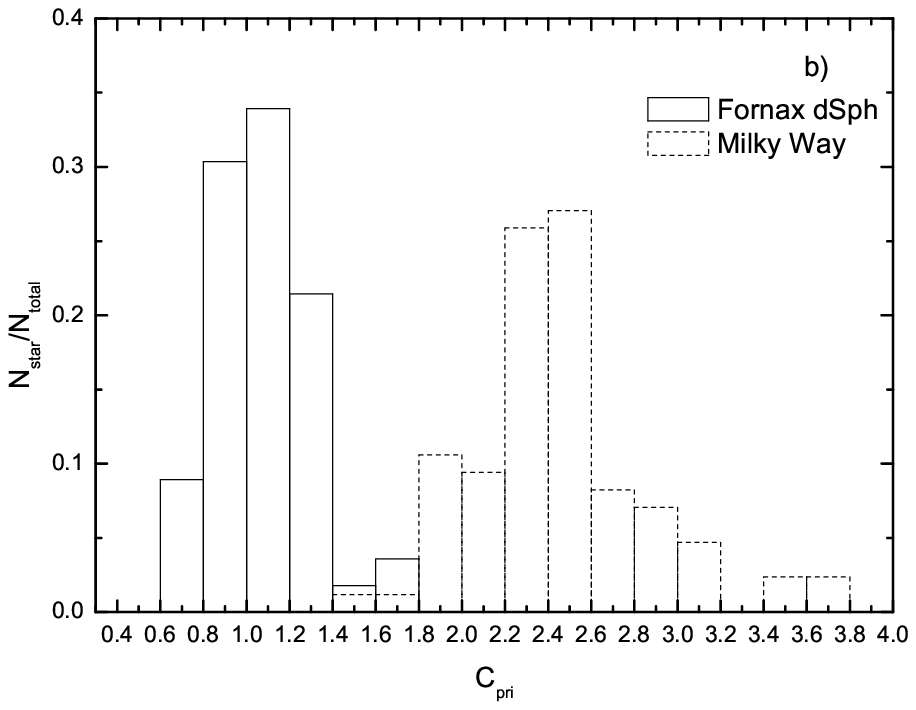}
 %\suppressfloats[t]
{Fig. 5 (continued).}
 %\label{appenfig}
\end{figure}

\begin{figure}[t]
 \centering
 \includegraphics[width=1\textwidth,height=0.6\textheight]{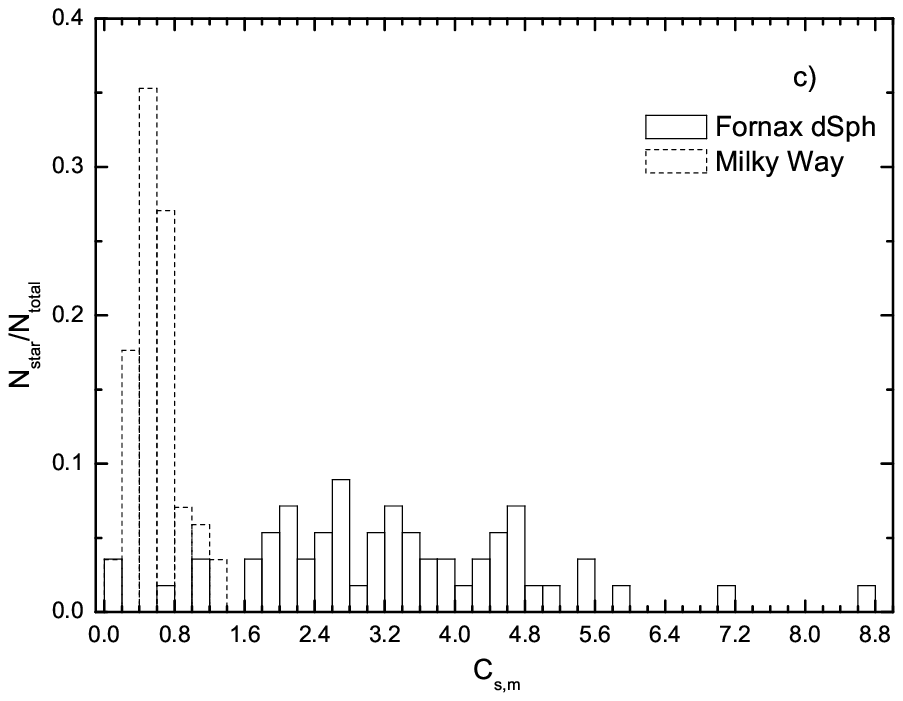}
 %\suppressfloats[t]
{Fig. 5 (continued).}
 %\label{appenfig}
\end{figure}

\begin{figure}[t]
 \centering
 \includegraphics[width=1\textwidth,height=0.6\textheight]{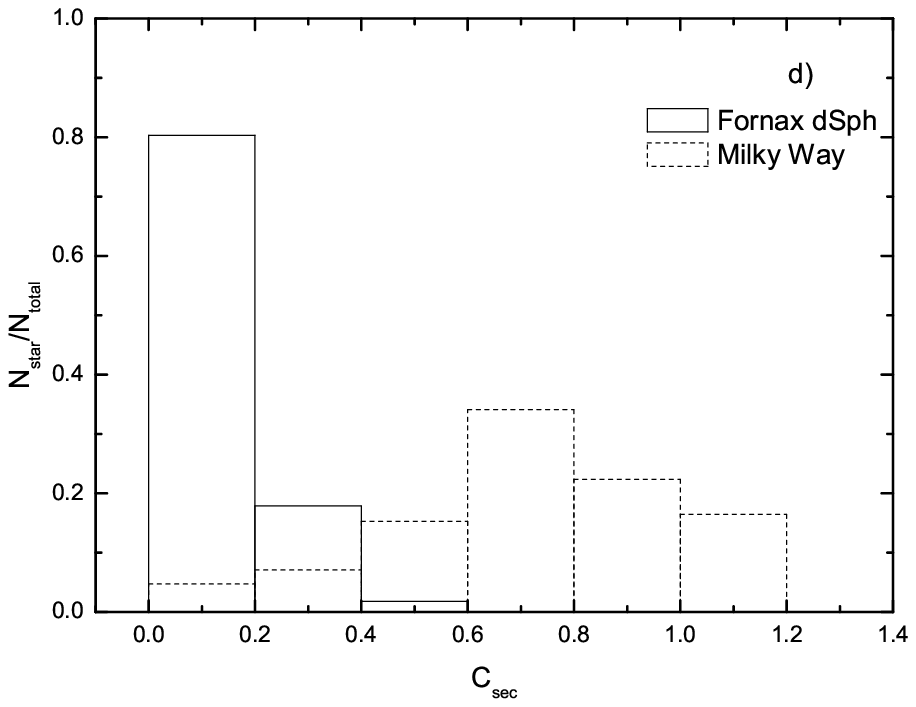}
 %\suppressfloats[t]
{Fig. 5 (continued).}
 %\label{appenfig}
\end{figure}

\begin{figure}[t]
 \centering
 \includegraphics[width=1\textwidth,height=0.6\textheight]{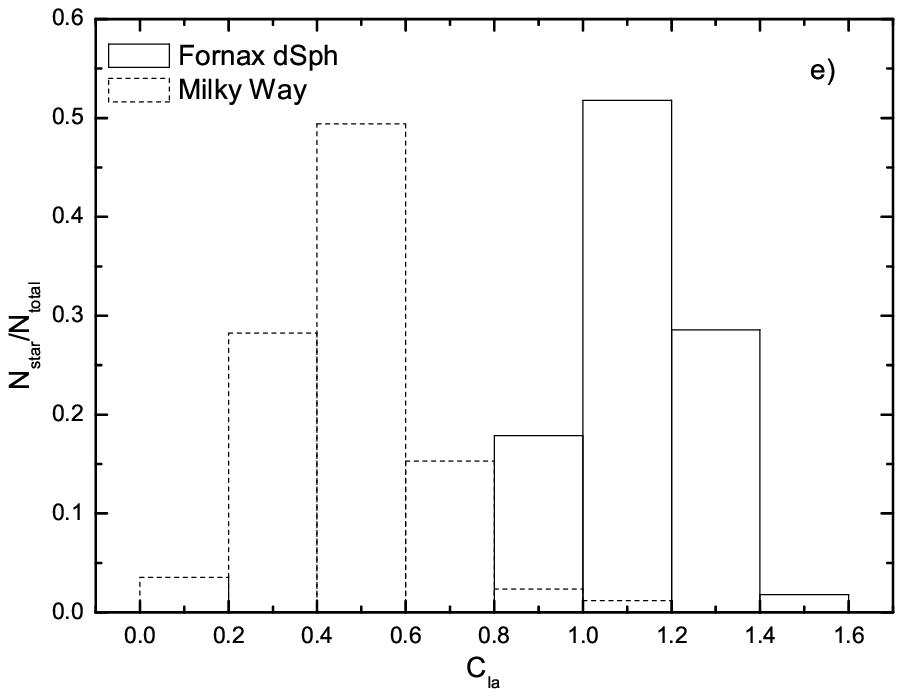}
 %\suppressfloats[t]
{Fig. 5 (continued).}
 %\label{appenfig}
\end{figure}

\begin{figure}[t]
 %\centering
 \includegraphics[width=1\textwidth,height=0.4\textheight]{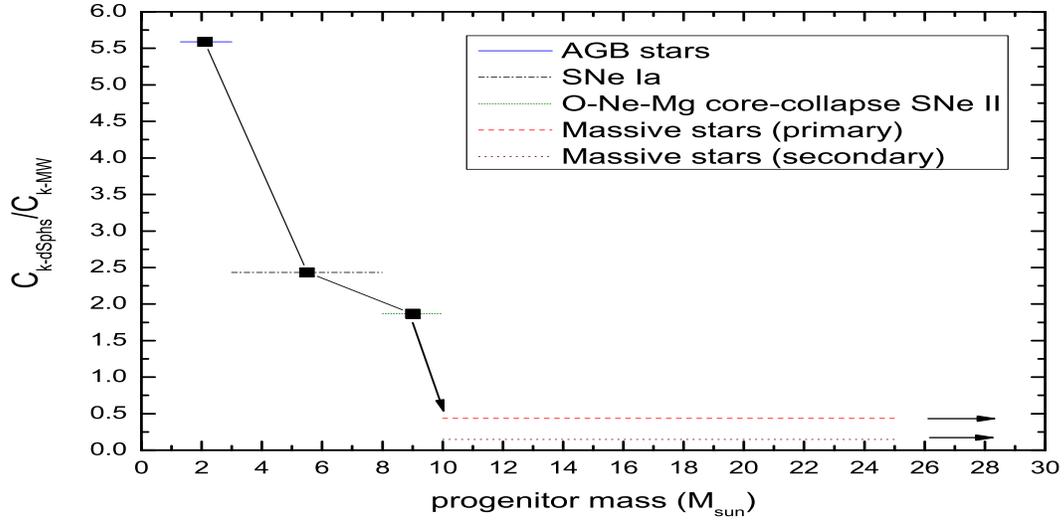}
 %\suppressfloats[t]
\caption{Ratios of the average component coefficient between the
dSph stars and the MW stars as a function of progenitor mass. The
solid line, short dotted line, dash line and dotted line represent
the progenitor mass rang responsible to main s-process, main
r-process, primary process and secondary process, respectively. The
dash dotted line is the progenitor mass rang of SNe Ia. The filled
square represents the average progenitor mass where the
corresponding process mainly occurs.}
  %\label{appenfig}
\end{figure}

\begin{figure}[b]
 \centering
 \includegraphics[width=1\textwidth,height=0.4\textheight]{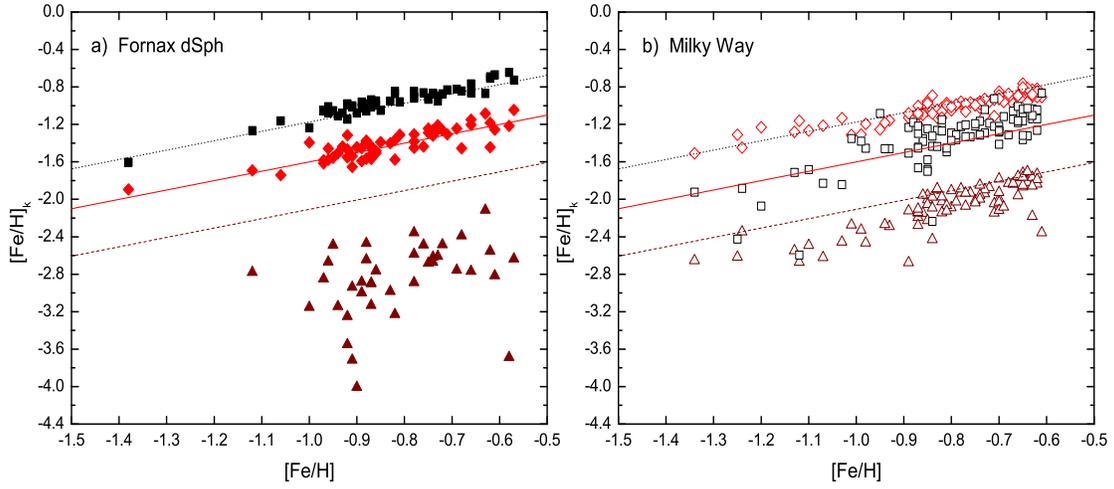}
 %\suppressfloats[t]
\caption{Component ratios of the individual process with various
metallicities for element Fe. The component ratios of the dSph stars
are shown in the left panel, in which the filled squares, filled
diamonds and filled triangles are the component ratios of SNe Ia,
primary-like yields and secondary-like yields, respectively. The
ratios of the MW stars are shown in the right panel, in which the
open squares, open diamonds and open triangles are the component
ratios of SNe Ia, primary-like yields and secondary-like yields,
respectively. In panels(a)and(b), solid lines, dash lines and dotted
lines represent the solar component ratios of primary-like yields,
secondary-like yields and SNe Ia, respectively.}
 %\label{appenfig}
\end{figure}

\begin{figure}
 %\centering
 \includegraphics[width=1\textwidth,height=0.4\textheight]{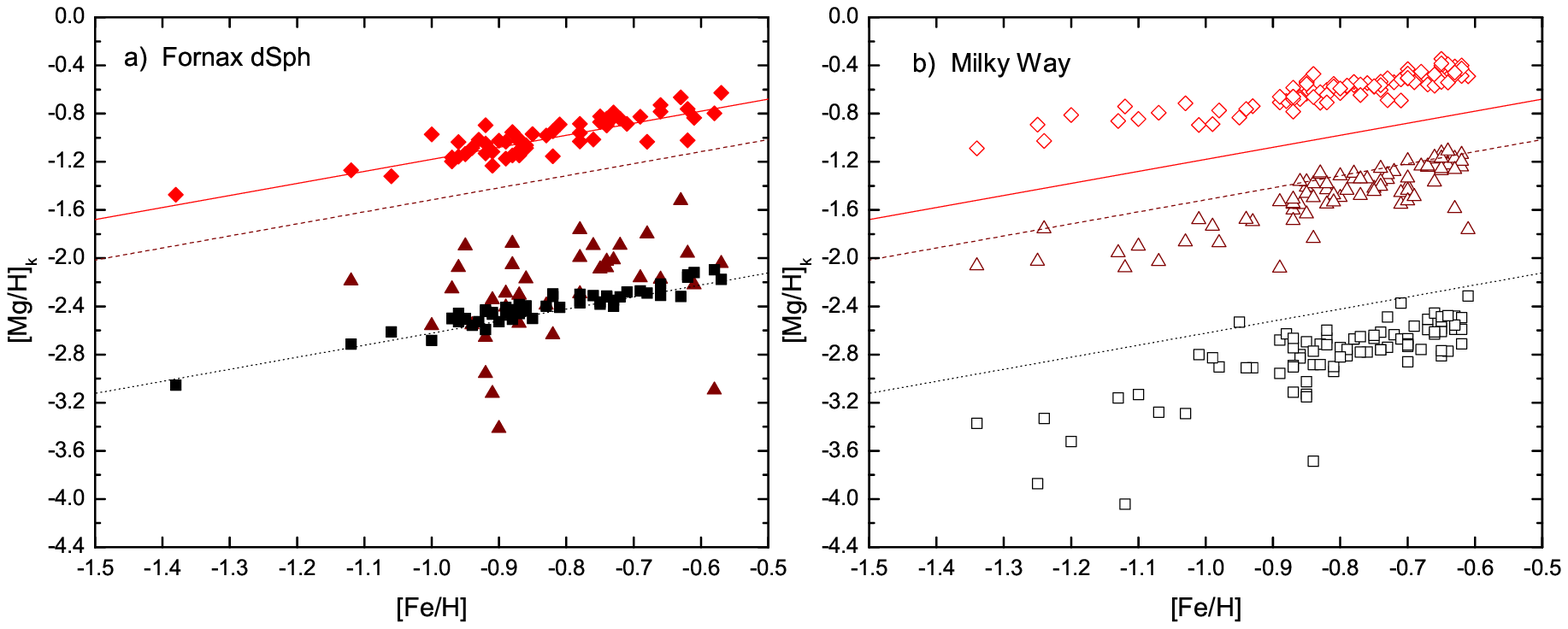}
 %\suppressfloats[t]
\caption{Component ratios of the individual process with various
metallicities for element Mg. The component ratios of the dSph stars
and the MW stars are shown in the left panel and right panel,
respectively. The symbols are the same as Fig. 7.}
 %\label{appenfig}
\end{figure}

\begin{figure}
 %\centering
 \includegraphics[width=1\textwidth,height=0.4\textheight]{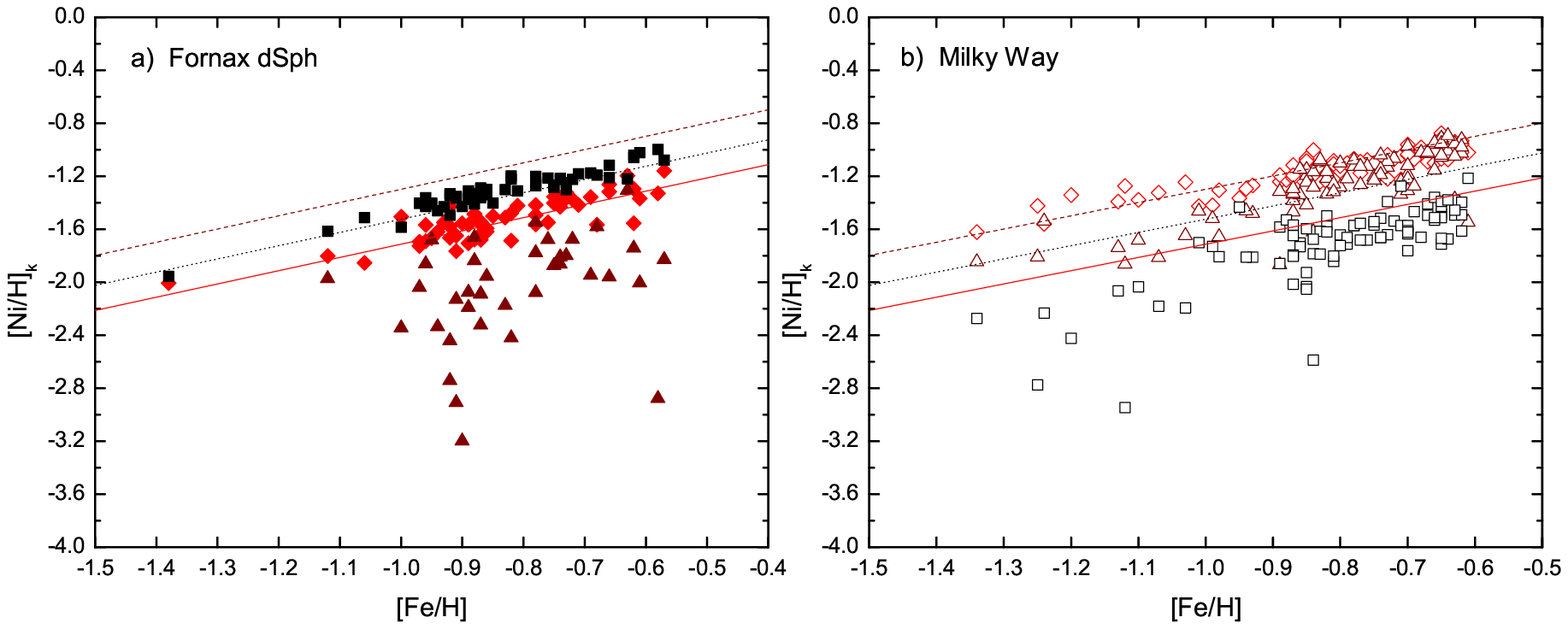}
 %\suppressfloats[t]
\caption{Component ratios of the individual process with various
metallicities for element Ni. The component ratios of the dSph stars
and the MW stars are shown in the left panel and right panel,
respectively. The symbols are the same as Fig. 7.}
 %\label{appenfig}
\end{figure}

\begin{figure}[b]
 \centering
 \includegraphics[width=1\textwidth,height=0.4\textheight]{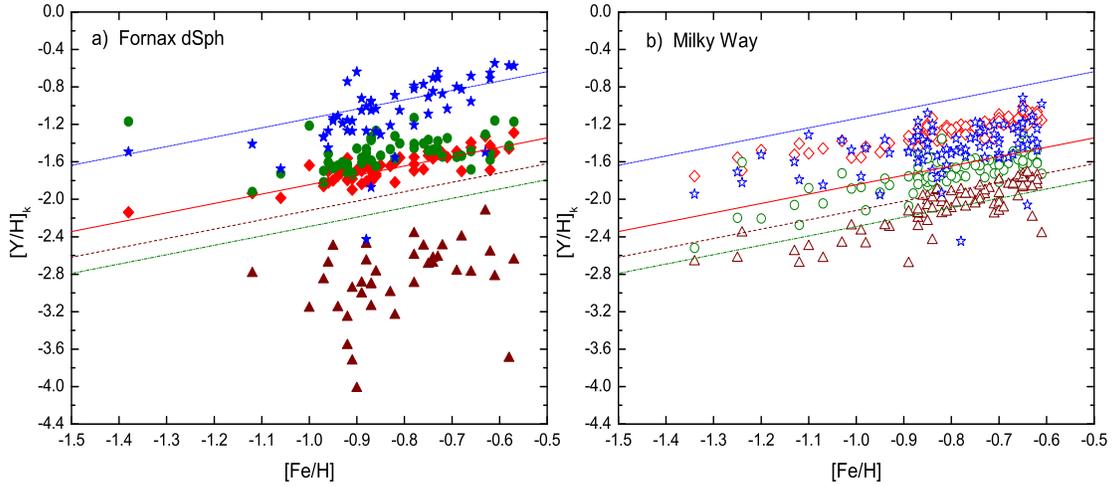}
 %\suppressfloats[t]
\caption{Component ratios of the individual process with various
metallicities for element Y. The component ratios of the dSph stars
are shown in the left panel, in which the filled circles, filled
diamonds, filled stars and filled triangles are the component ratios
of main r-process, weak r-process, main s-process and weak
s-process, respectively. The ratios of the MW stars are shown in the
right panel, in which the open circles, open diamonds, open stars
and open triangles are responsible to main r-process, weak
r-process, main s-process and weak s-process, respectively. In
panels(a)and(b), dash dotted lines, solid lines, short dotted lines
and dash lines represent the solar component ratios of main r-, weak
r-, main s- and weak s-process, respectively.}
 %\label{appenfig}
\end{figure}

\begin{figure}[b]
 \centering
 \includegraphics[width=1\textwidth,height=0.4\textheight]{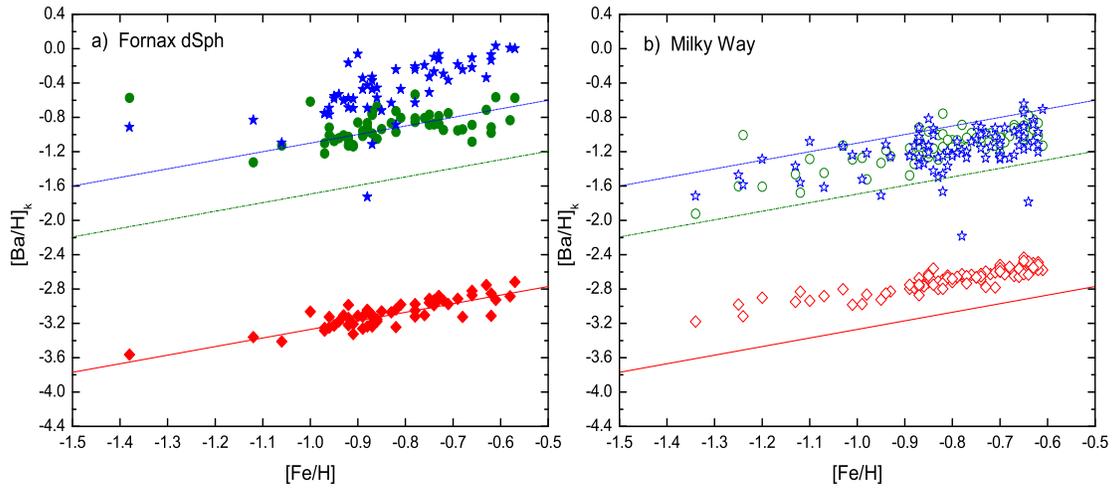}
 %\suppressfloats[t]
\caption{Component ratios of the individual process with various
metallicities for element Ba. The component ratios of the dSph stars
and the MW stars are shown in the left panel and right panel,
respectively. The symbols are similar with Fig. 10.}
 %\label{appenfig}
\end{figure}

\clearpage

\begin{table}
\begin{center}
\caption{The five component coefficients, $\chi^{2}$ and
$K-K{_free}$ for the Fornax dSph stars.\label{tbl-1}}

\begin{tabular*}{458pt}{@{\extracolsep\fill}lrrrrrrrc}

\tableline\tableline Star  &  [Fe/H]  & $C_{r,m}$ & $C_{pri}$ &
$C_{s,m}$ &$C_{sec}$  &  $C_{Ia}$ & $\chi^2$
& $K-K_{free}$ \\
\tableline
BL038   &   -0.88   &   5.10    &   1.28    &   3.55    &   0.22    &   1.13    &   1.88    &   8   \\
BL079   &   -0.57   &   4.90    &   1.33    &   4.73    &   0.11    &   1.04    &   0.49    &   8   \\
BL081   &   -0.62   &   2.15    &   0.60    &   3.87    &   0.15    &   1.22    &   2.65    &   8   \\
BL091   &   -0.96   &   4.43    &   0.96    &   1.96    &   0.25    &   1.33    &   0.62    &   8   \\
BL092   &   -0.95   &   3.71    &   0.99    &   3.13    &   0.00    &   1.18    &   1.97    &   8   \\
BL096   &   -0.75   &   4.74    &   1.15    &   2.20    &   0.00    &   1.02    &   0.75    &   8   \\
BL097   &   -0.92   &   3.86    &   1.13    &   2.16    &   0.03    &   1.30    &   1.22    &   8   \\
BL104   &   -0.96   &   5.45    &   1.27    &   2.34    &   0.00    &   1.14    &   2.40    &   8   \\
BL113   &   -0.75   &   3.62    &   1.28    &   3.34    &   0.15    &   0.98    &   0.62    &   8   \\
BL123   &   -0.97   &   3.66    &   0.90    &   2.09    &   0.17    &   1.23    &   1.95    &   8   \\
BL125   &   -0.73   &   4.37    &   1.20    &   5.11    &   0.00    &   0.90    &   1.15    &   8   \\
BL135   &   -0.95   &   3.86    &   0.99    &   2.94    &   0.37    &   1.16    &   2.03    &   8   \\
BL140   &   -0.87   &   6.41    &   0.88    &   3.16    &   0.07    &   1.28    &   1.97    &   8   \\
BL141   &   -0.82   &   3.50    &   0.70    &   1.08    &   0.00    &   1.33    &   0.86    &   8   \\
BL146   &   -0.92   &   3.06    &   0.93    &   2.78    &   0.00    &   1.26    &   3.21    &   8   \\
BL147   &   -1.38   &   31.69   &   1.22    &   3.69    &   0.00    &   0.89    &   1.96    &   8   \\
BL148   &   -0.63   &   4.09    &   1.40    &   2.47    &   0.42    &   0.86    &   1.54    &   8   \\
BL149   &   -0.91   &   2.92    &   0.94    &   2.67    &   0.02    &   1.17    &   1.35    &   8   \\
BL155   &   -0.74   &   3.57    &   1.20    &   5.53    &   0.17    &   1.12    &   0.90    &   8   \\
BL158   &   -0.87   &   6.69    &   0.80    &   4.37    &   0.12    &   1.26    &   2.06    &   8   \\
BL160   &   -0.93   &   4.13    &   1.24    &   2.67    &   0.00    &   1.06    &   1.29    &   8   \\
BL163   &   -0.78   &   4.60    &   1.00    &   4.37    &   0.20    &   1.12    &   1.09    &   8   \\
BL166   &   -0.89   &   3.69    &   1.09    &   4.44    &   0.13    &   1.10    &   0.84    &   8   \\
BL168   &   -0.88   &   4.61    &   1.27    &   1.95    &   0.33    &   0.99    &   1.24    &   8   \\
BL173   &   -0.86   &   3.19    &   0.95    &   2.45    &   0.16    &   1.11    &   0.96    &   8   \\
BL180   &   -0.9    &   5.42    &   1.14    &   8.71    &   0.01    &   0.99    &   0.40    &   8   \\
BL185   &   -0.73   &   3.54    &   1.32    &   5.85    &   0.17    &   1.01    &   0.98    &   8   \\
BL196   &   -1.06   &   4.22    &   0.83    &   1.16    &   0.00    &   1.18    &   1.54    &   8   \\
BL197   &   -0.89   &   4.20    &   0.79    &   3.31    &   0.10    &   1.27    &   2.12    &   8   \\
BL203   &   -0.83   &   4.54    &   1.07    &   1.99    &   0.09    &   1.13    &   0.32    &   8   \\
BL204   &   -1  &   11.94   &   1.61    &   0.00    &   0.09    &   0.87    &   0.42    &   8   \\
BL205   &   -0.69   &   2.71    &   1.11    &   4.06    &   0.11    &   1.10    &   0.34    &   8   \\
BL208   &   -0.66   &   1.86    &   1.29    &   3.48    &   0.00    &   0.94    &   1.21    &   8   \\
BL210   &   -0.76   &   4.44    &   0.84    &   4.66    &   0.24    &   1.18    &   2.14    &   8   \\
BL211   &   -0.66   &   2.92    &   1.14    &   4.52    &   0.10    &   1.17    &   2.25    &   8   \\
BL213   &   -0.94   &   3.90    &   1.08    &   3.23    &   0.08    &   1.01    &   1.33    &   8   \\
BL216   &   -0.78   &   4.16    &   1.19    &   4.70    &   0.34    &   1.08    &   1.18    &   8   \\
BL218   &   -0.62   &   2.50    &   1.09    &   4.47    &   0.00    &   1.27    &   1.24    &   8   \\
BL221   &   -0.86   &   7.61    &   0.90    &   3.19    &   0.00    &   1.22    &   0.79    &   8   \\
\tableline
\end{tabular*}
\clearpage
\end{center}
\end{table}

\begin{table}
\begin{center}
\begin{tabular*}{458pt}{@{\extracolsep\fill}lrrrrrrrc}

\tableline\tableline Star  &  [Fe/H]  & $C_{r,m}$ & $C_{pri}$ &
$C_{s,m}$ &$C_{sec}$   &  $C_{Ia}$ & $\chi^2$
& $K-K_{free}$  \\
\tableline
BL227   &   -0.87   &   6.22    &   1.09    &   3.98    &   0.12    &   1.08    &   1.33    &   8   \\
BL228   &   -0.88   &   4.04    &   0.82    &   0.18    &   0.00    &   1.19    &   3.75    &   8   \\
BL229   &   -0.71   &   4.51    &   1.01    &   2.78    &   0.00    &   1.13    &   0.82    &   8   \\
BL233   &   -0.68   &   2.73    &   0.67    &   3.43    &   0.25    &   1.03    &   2.08    &   8   \\
BL242   &   -1.12   &   3.09    &   1.07    &   2.45    &   0.28    &   1.07    &   2.41    &   8   \\
BL247   &   -0.82   &   6.10    &   1.13    &   4.77    &   0.05    &   1.40    &   3.54    &   8   \\
BL253   &   -0.74   &   3.64    &   1.05    &   3.74    &   0.15    &   1.10    &   1.02    &   8   \\
BL257   &   -0.58   &   2.76    &   0.92    &   4.87    &   0.01    &   1.28    &   1.18    &   8   \\
BL258   &   -0.61   &   5.50    &   0.90    &   5.53    &   0.08    &   1.30    &   0.44    &   8   \\
BL260   &   -0.87   &   3.77    &   1.05    &   0.72    &   0.00    &   1.11    &   1.31    &   8   \\
BL261   &   -0.85   &   4.04    &   1.15    &   1.70    &   0.00    &   0.94    &   1.48    &   8   \\
BL267   &   -0.72   &   2.93    &   1.14    &   3.36    &   0.22    &   1.05    &   0.56    &   8   \\
BL269   &   -0.81   &   4.99    &   1.26    &   2.75    &   0.00    &   1.06    &   1.44    &   8   \\
BL300   &   -0.92   &   3.50    &   1.60    &   7.19    &   0.06    &   0.89    &   1.78    &   8   \\
BL304   &   -0.97   &   2.77    &   0.97    &   2.06    &   0.00    &   1.24    &   1.53    &   8   \\
BL311   &   -0.78   &   8.69    &   0.85    &   1.78    &   0.10    &   1.27    &   0.80    &   8   \\
BL323   &   -0.91   &   3.04    &   0.72    &   2.08    &   0.12    &   1.21    &   2.23    &   8   \\
\tableline
\end{tabular*}
\clearpage
\end{center}
\end{table}

\clearpage

\begin{table}
\begin{center}
\caption{The five component coefficients, $\chi^{2}$ and
$K-K{_free}$ for the MW stars.\label{tbl-2}}

\begin{tabular*}{458pt}{@{\extracolsep\fill}lrrrrrrrc}

\tableline\tableline Star  &  [Fe/H]  & $C_{r,m}$ & $C_{pri}$ &
$C_{s,m}$ &$C_{sec}$  &  $C_{Ia}$ & $\chi^2$
& $K-K_{free}$ \\
\tableline
HD 245  &   -0.78   &   3.86    &   2.50    &   0.05    &   0.90    &   0.54    &   0.97    &   6   \\
HD 3546 &   -0.63   &   2.48    &   2.52    &   0.45    &   0.36    &   0.59    &   0.20    &   6   \\
HD 6833 &   -0.89   &   1.28    &   2.30    &   0.56    &   0.21    &   0.68    &   0.78    &   6   \\
HD 64606    &   -0.82   &   2.33    &   2.50    &   0.29    &   0.80    &   0.60    &   0.61    &   6   \\
HD 105755   &   -0.65   &   1.70    &   3.06    &   1.28    &   0.00    &   0.49    &   0.76    &   6   \\
HD 108076   &   -0.85   &   1.92    &   2.49    &   1.37    &   0.90    &   0.60    &   1.50    &   6   \\
HD 127243   &   -0.65   &   1.94    &   2.50    &   0.71    &   1.00    &   0.46    &   0.34    &   6   \\
HD 166161   &   -1.20   &   1.93    &   3.71    &   1.03    &   0.00    &   0.20    &   1.45    &   6   \\
HD 204155   &   -0.78   &   2.12    &   2.65    &   0.50    &   1.00    &   0.54    &   2.41    &   6   \\
HD 208906   &   -0.71   &   2.36    &   1.58    &   0.77    &   0.47    &   0.91    &   0.82    &   6   \\
HD 224930   &   -0.85   &   2.12    &   3.08    &   0.67    &   1.00    &   0.28    &   2.17    &   6   \\
HD 165908   &   -0.61   &   1.49    &   2.00    &   1.01    &   0.23    &   0.83    &   1.22    &   5   \\
HD 221377   &   -0.88   &   1.71    &   2.31    &   0.71    &   0.00    &   0.75    &   1.50    &   5   \\
HIP 3185    &   -0.65   &   4.32    &   2.68    &   0.38    &   1.09    &   0.29    &   0.53    &   8   \\
HIP 5122    &   -0.62   &   2.80    &   2.05    &   0.40    &   0.99    &   0.56    &   0.49    &   8   \\
HIP 5336    &   -0.86   &   2.30    &   2.39    &   0.42    &   1.03    &   0.48    &   0.55    &   8   \\
HIP 6159    &   -0.67   &   2.97    &   2.20    &   0.64    &   0.87    &   0.50    &   0.78    &   8   \\
HIP 7961    &   -0.64   &   2.08    &   2.69    &   0.40    &   0.85    &   0.31    &   0.42    &   8   \\
HIP 12579   &   -0.80   &   2.69    &   2.24    &   0.47    &   0.66    &   0.48    &   0.75    &   8   \\
HIP 13366   &   -0.70   &   2.41    &   2.55    &   0.33    &   0.65    &   0.39    &   0.60    &   8   \\
HIP 15405   &   -0.73   &   1.80    &   1.67    &   0.70    &   0.79    &   0.73    &   0.24    &   8   \\
HIP 17147   &   -0.87   &   2.74    &   2.93    &   0.78    &   0.61    &   0.24    &   1.08    &   8   \\
HIP 17666   &   -1.03   &   3.96    &   3.14    &   0.99    &   0.48    &   0.23    &   1.00    &   8   \\
HIP 22060   &   -0.63   &   2.43    &   2.14    &   0.79    &   0.94    &   0.46    &   1.47    &   8   \\
HIP 29269   &   -0.68   &   2.27    &   2.56    &   0.32    &   0.91    &   0.35    &   0.33    &   8   \\
HIP 39893   &   -0.84   &   3.78    &   2.50    &   0.33    &   0.94    &   0.38    &   0.74    &   8   \\
HIP 40613   &   -0.62   &   2.02    &   2.52    &   0.33    &   0.78    &   0.34    &   0.80    &   8   \\
HIP 44075   &   -0.86   &   2.27    &   2.30    &   0.69    &   0.66    &   0.45    &   0.66    &   8   \\
HIP 44347   &   -0.85   &   3.80    &   3.14    &   0.66    &   0.54    &   0.22    &   0.76    &   8   \\
HIP 52673   &   -0.66   &   2.66    &   2.31    &   0.65    &   1.00    &   0.45    &   0.75    &   8   \\
HIP 59233   &   -0.83   &   1.70    &   2.44    &   0.27    &   1.10    &   0.37    &   0.76    &   8   \\
HIP 60268   &   -0.72   &   2.69    &   2.30    &   0.48    &   0.90    &   0.51    &   0.36    &   8   \\
HIP 62240   &   -0.83   &   3.30    &   2.00    &   0.46    &   1.13    &   0.55    &   1.84    &   8   \\
HIP 64426   &   -0.71   &   1.89    &   2.37    &   0.61    &   0.59    &   0.46    &   0.69    &   8   \\
HIP 70520   &   -0.62   &   1.97    &   2.37    &   0.56    &   0.87    &   0.45    &   0.40    &   8   \\
HIP 72803   &   -0.73   &   2.73    &   2.53    &   0.36    &   0.87    &   0.41    &   1.02    &   8   \\
HIP 74033   &   -0.85   &   2.39    &   2.99    &   0.55    &   0.80    &   0.21    &   0.92    &   8   \\
HIP 74067   &   -0.75   &   2.99    &   2.26    &   0.62    &   0.70    &   0.51    &   0.82    &   8   \\
\tableline
\end{tabular*}
\clearpage
\end{center}
\end{table}

\begin{table}
\begin{center}
\begin{tabular*}{458pt}{@{\extracolsep\fill}lrrrrrrrc}

\tableline\tableline Star  &  [Fe/H]  & $C_{r,m}$ & $C_{pri}$ &
$C_{s,m}$ &$C_{sec}$   &  $C_{Ia}$ & $\chi^2$
& $K-K_{free}$  \\
\tableline
HIP 85757   &   -0.70   &   1.50    &   2.82    &   0.53    &   0.76    &   0.29    &   0.65    &   8   \\
HIP 86013   &   -0.70   &   2.74    &   2.35    &   0.48    &   0.60    &   0.45    &   0.92    &   8   \\
HIP 88039   &   -0.81   &   1.96    &   2.74    &   0.52    &   0.71    &   0.31    &   1.08    &   8   \\
HIP 88166   &   -0.76   &   2.95    &   2.48    &   0.60    &   0.86    &   0.40    &   1.19    &   8   \\
HIP 98532   &   -1.13   &   2.29    &   2.82    &   0.73    &   0.49    &   0.39    &   0.88    &   8   \\
HIP 112811  &   -0.70   &   1.88    &   2.58    &   0.37    &   0.49    &   0.41    &   1.06    &   8   \\
HIP 113514  &   -0.63   &   3.17    &   2.24    &   0.70    &   0.76    &   0.50    &   0.70    &   8   \\
HIP 117029  &   -0.77   &   2.02    &   2.46    &   0.47    &   0.88    &   0.41    &   0.64    &   8   \\
HIP 4039    &   -1.24   &   8.45    &   2.47    &   0.57    &   1.00    &   0.34    &   0.81    &   6   \\
HIP 4544    &   -0.87   &   4.41    &   2.45    &   0.51    &   0.68    &   0.40    &   1.79    &   7   \\
HIP 10652   &   -0.67   &   2.75    &   1.96    &   0.34    &   0.89    &   0.61    &   0.39    &   7   \\
HIP 15126   &   -0.82   &   3.30    &   2.27    &   0.47    &   0.61    &   0.53    &   0.44    &   6   \\
HIP 26452   &   -0.89   &   2.66    &   2.57    &   0.55    &   0.75    &   0.36    &   0.97    &   7   \\
HIP 27128   &   -0.81   &   2.76    &   2.60    &   0.35    &   0.62    &   0.34    &   1.24    &   7   \\
HIP 36849   &   -0.77   &   2.68    &   2.02    &   0.40    &   0.64    &   0.55    &   0.78    &   7   \\
HIP 38769   &   -0.79   &   2.05    &   2.49    &   0.47    &   0.74    &   0.40    &   0.52    &   7   \\
HIP 58843   &   -0.79   &   2.62    &   2.52    &   0.64    &   0.74    &   0.45    &   0.86    &   8   \\
HIP 59750   &   -0.74   &   2.19    &   2.43    &   0.69    &   0.78    &   0.40    &   1.24    &   6   \\
HIP 70681   &   -1.10   &   3.23    &   2.72    &   1.32    &   0.52    &   0.39    &   1.21    &   7   \\
HIP 85373   &   -0.82   &   5.72    &   1.96    &   0.18    &   0.90    &   0.70    &   2.88    &   7   \\
HIP 104659  &   -1.07   &   2.07    &   2.87    &   0.36    &   0.36    &   0.26    &   1.12    &   6   \\
HIP 110291  &   -0.93   &   2.36    &   2.36    &   0.59    &   0.56    &   0.44    &   1.24    &   7   \\
HIP 14086   &   -0.65   &   2.81    &   2.80    &   0.53    &   0.78    &   0.32    &   0.84    &   8   \\
HIP 10449   &   -0.87   &   3.22    &   2.34    &   0.88    &   0.50    &   0.56    &   0.38    &   7   \\
HIP 28671   &   -1.01   &   2.60    &   1.98    &   0.74    &   0.70    &   0.68    &   0.92    &   6   \\
HIP 55592   &   -0.94   &   2.83    &   2.28    &   0.84    &   0.60    &   0.45    &   0.83    &   7   \\
HIP 62882   &   -0.98   &   1.42    &   2.43    &   0.73    &   0.42    &   0.50    &   0.55    &   8   \\
HIP 78640   &   -1.34   &   1.29    &   2.70    &   0.53    &   0.62    &   0.39    &   0.39    &   5   \\
HIP 80837   &   -0.80   &   2.21    &   2.45    &   0.45    &   1.00    &   0.40    &   0.68    &   8   \\
HIP 86321   &   -0.87   &   2.44    &   1.85    &   1.12    &   0.69    &   0.67    &   0.37    &   6   \\
HIP 94449   &   -1.12   &   1.37    &   3.63    &   0.46    &   0.36    &   0.05    &   0.58    &   7   \\
HIP 100568  &   -0.99   &   2.58    &   1.92    &   0.37    &   0.59    &   0.61    &   0.39    &   8   \\
HIP 111549  &   -0.95   &   2.05    &   1.98    &   0.22    &   0.00    &   1.10    &   1.31    &   6   \\
HIP 5163    &   -0.74   &   2.51    &   2.08    &   0.85    &   1.00    &   0.56    &   0.59    &   7   \\
HIP 8720    &   -0.74   &   2.47    &   2.28    &   0.48    &   0.71    &   0.40    &   1.13    &   8   \\
HIP 10711   &   -0.69   &   1.97    &   2.14    &   0.74    &   0.52    &   0.56    &   0.70    &   7   \\
HIP 21306   &   -0.65   &   2.10    &   2.11    &   1.01    &   0.81    &   0.56    &   0.71    &   8   \\
HIP 30990   &   -0.87   &   2.10    &   2.45    &   0.41    &   0.75    &   0.39    &   0.67    &   7   \\
HIP 42734   &   -0.70   &   2.29    &   2.38    &   0.34    &   1.06    &   0.40    &   0.22    &   8   \\
HIP 80162   &   -0.66   &   2.52    &   1.87    &   0.57    &   1.03    &   0.67    &   1.21    &   8   \\
\tableline
\end{tabular*}
\clearpage
\end{center}
\end{table}

\begin{table}
\begin{center}
\begin{tabular*}{458pt}{@{\extracolsep\fill}lrrrrrrrc}

\tableline\tableline Star  &  [Fe/H]  & $C_{r,m}$ & $C_{pri}$ &
$C_{s,m}$ &$C_{sec}$   &  $C_{Ia}$ & $\chi^2$
& $K-K_{free}$  \\
\tableline
HIP 20298   &   -0.84   &   2.10    &   3.54    &   1.02    &   0.33    &   0.06    &   1.21    &   8   \\
HIP 23922   &   -0.66   &   2.38    &   2.32    &   0.43    &   0.64    &   0.47    &   0.99    &   8   \\
HIP 26617   &   -0.75   &   2.54    &   2.28    &   0.89    &   0.66    &   0.54    &   0.58    &   8   \\
HIP 43595   &   -0.84   &   2.35    &   2.28    &   0.86    &   0.72    &   0.49    &   0.69    &   8   \\
HIP 48209   &   -0.65   &   2.24    &   1.96    &   0.69    &   0.99    &   0.61    &   0.54    &   8   \\
HIP 51477   &   -1.25   &   2.20    &   3.46    &   0.76    &   0.55    &   0.10    &   1.77    &   8   \\
HIP 60331   &   -0.64   &   3.61    &   1.90    &   0.09    &   1.11    &   0.61    &   1.63    &   7   \\
\tableline
\end{tabular*}
\clearpage
\end{center}
\end{table}

\end{document}